\documentclass[aps,prd,twocolumn,nofootinbib,superscriptaddress]{revtex4-2}

\usepackage{amsmath,amssymb,bm}

\begin{document}

\title{Detector Resolution and Observable Infrared Memory in QED}

\author{Takeshi Fukuyama}
\affiliation{Research Center for Nuclear Physics (RCNP), Osaka University,
Ibaraki, Osaka 567-0047, Japan}

\begin{abstract}
Infrared divergences in QED cancel in inclusive observables, while the physical detector resolution $\omega_{\max}$ remains. We relate this finite-resolution description to electromagnetic memory through the reduced density matrix. For unresolved soft-photon clouds, the logarithmic resolution flow of hard-sector coherence is determined by one half of the squared distance between the corresponding memory records. For infrared-safe dressed asymptotic states, however, the leading memory contribution is absorbed into the dressing and the resolution flow vanishes in the strict infrared limit. The contrast between these two behaviors provides an operational diagnostic of how electromagnetic memory is encoded in the asymptotic quantum state.
\end{abstract}
\maketitle

\section{Introduction}

Infrared divergences in quantum electrodynamics arise from the universal
coupling of charged particles to photons of arbitrarily small energy.
In the conventional formulation, individual Fock-space amplitudes are
infrared divergent.  Physical observables are obtained only after virtual
soft-photon corrections are combined with real soft-photon emission below
a finite experimental resolution scale.

The standard Bloch--Nordsieck and Kinoshita--Lee--Nauenberg mechanisms
show that the unphysical infrared regulator cancels in inclusive
observables \cite{BlochNordsieck, Kinoshita, LeeNauenberg}.  This result is usually interpreted as the resolution of the
infrared problem.  However, the inclusive observable still depends on the
detector resolution scale. This scale is the maximum energy $\omega_{\max}$ of unresolved soft photons \cite{LL}.
The key step of the present work is to combine the operational
interpretation of $\omega_{\max}$, familiar from the conventional
infrared treatment of finite detector resolution, with the modern
identification of the leading soft sector with electromagnetic
memory.  In this way, the conventional resolution scale acquires a
direct interpretation as a scale controlling the observable
distinguishability of electromagnetic memory records.

The purpose of this Letter is to formulate this connection quantitatively.  In the density-matrix description, $\omega_{\max}$ specifies the boundary between observed and unresolved soft modes.  Tracing over the latter gives a resolution-dependent overlap between the soft-photon clouds associated with different hard scattering branches.  The coefficient of the leading soft factor defines the corresponding electromagnetic memory record, allowing us to introduce a distance between two such records.

For the unresolved soft-cloud description, we show that the logarithmic variation of hard-sector coherence with detector resolution is determined by one half of the squared memory distance.  We then examine the same resolution dependence for infrared-safe dressed asymptotic states.  In this case the leading $1/\omega$ memory contribution is absorbed into the asymptotic dressing and cancels from the interference factor, leaving only the subleading dressing dependence.  Consequently, the corresponding resolution-flow rate vanishes in the strict infrared limit.

The contrast between these two infrared behaviors is the main result of the present work.  Rather than introducing a new infrared scale, we use the conventional detector resolution $\omega_{\max}$ as a variable operational parameter.  Its variation probes how soft information is partitioned between unresolved radiation and asymptotic dressing, thereby connecting the traditional finite-resolution formulation of infrared QED with the modern descriptions of electromagnetic memory, dressed states, and quantum coherence.

\section{Infrared cancellation and the surviving resolution scale}

Let $\lambda$ denote an auxiliary photon mass introduced as an infrared
regulator.  Virtual soft-photon corrections contain logarithmic terms of
the form
\begin{equation}
\delta_{\rm virt}^{\rm IR}
=
-\frac{2\alpha}{\pi}
\ln\frac{{\bf q}^2}{m^2}
\ln\frac{m}{\lambda}
+\cdots .
\end{equation}
On the other hand, the real emission of unresolved soft photons with
energy below the detector resolution scale $\omega_{\max}$ gives
\begin{equation}
d\sigma_{\rm soft}
=
\frac{2\alpha}{\pi}
\left[
\ln\frac{{\bf q}^2}{m^2}
\ln\frac{\omega_{\max}}{\lambda}
+\cdots
\right]
d\sigma_{\rm el}.
\end{equation}
Thus the infrared-sensitive combination is
\begin{equation}
\ln\frac{\omega_{\max}}{\lambda}
-
\ln\frac{m}{\lambda}
=
\ln\frac{\omega_{\max}}{m}.
\end{equation}
The auxiliary photon mass $\lambda$ cancels, while the detector resolution
scale $\omega_{\max}$ remains.

Therefore the infrared-safe observable is not independent of infrared
resolution.  Rather, it is a quantity defined at a finite resolution,
\begin{equation}
d\sigma_{\rm incl}
=
d\sigma_{\rm incl}(\omega_{\max}).
\end{equation}
The conventional statement that infrared divergences cancel should
therefore be understood more precisely: the unphysical regulator
dependence cancels, but the physical resolution dependence survives.
\section{Reduced Density Matrix and Soft-Sector Overlap}

The same structure has a natural interpretation in the
density-matrix language. Consider an asymptotic hard
state prepared as a superposition of charged momentum
configurations. In QED the corresponding physical state
is not a bare hard state, but a dressed state,

\begin{equation}
|\Psi\rangle
=
\sum_i c_i
|p_i\rangle_{\rm hard}
\otimes
|\gamma_i\rangle_{\rm soft},
\end{equation}

where $|\gamma_i\rangle_{\rm soft}$ denotes the coherent
soft-photon cloud associated with the hard configuration
$p_i$. At this stage we use ``dressed state'' in the general sense of a hard configuration accompanied by its coherent soft-photon cloud; the additional constraints required for a superselection-compatible infrared-safe asymptotic dressing will be imposed separately in Sec.~IV.

The infrared quantum-information viewpoint suggests that unresolved soft photons should be treated as unobserved degrees of freedom [5--8].  Related formulations of infrared-finite density matrices and the loss of quantum coherence induced by tracing over unresolved radiation were developed in Refs.~\cite{GomezLetschkaZell2018a,GomezLetschkaZell2018b}, where the role of the detector energy resolution as a physical infrared scale was also emphasized.

We propose to identify the detector resolution scale
$\omega_{\max}$, which appears in inclusive observables,
with the boundary separating observed and unobserved
soft modes.
Accordingly, we define
\begin{equation}
\rho_{\rm hard}
=
{\rm Tr}_{\omega<\omega_{\max}}
\left(|\Psi\rangle\langle\Psi|\right).
\end{equation}

It can be written as

\begin{equation}
\rho_{\rm hard}
=
\sum_{i,j}
c_i c_j^*
D_{ij}(\omega_{\max})
|p_i\rangle\langle p_j|,
\end{equation}

where

\begin{equation}
D_{ij}(\omega_{\max})
=
{}_{\rm soft}\langle\gamma_j|\gamma_i\rangle_{\rm soft}
\end{equation}

is the overlap of the unresolved soft-photon clouds.
For coherent-state dressings restricted to the unresolved
soft sector,
\begin{equation}
|\gamma_i\rangle_{\rm soft}
=
\exp\!\left[
\int_{\lambda<\omega<\omega_{\max}}
\frac{d^3k}{2\omega_k}
\left(f_i(k)a^\dagger(k)-f_i^*(k)a(k)\right)
\right]|0\rangle ,
\end{equation}

the standard coherent-state overlap gives
\begin{align}
{}_{\rm soft}\langle\gamma_j|\gamma_i\rangle_{\rm soft}
&=
\exp\Biggl[
-\frac12
\int_{\lambda<\omega<\omega_{\max}}
\frac{d^3k}{2\omega_k}
|f_i(k)-f_j(k)|^2
\nonumber\\
&\qquad\qquad
+i\Phi_{ij}
\Biggr],
\end{align}
where $\Phi_{ij}$ denotes an unimportant relative phase.

For the Faddeev--Kulish dressing one has \cite{Chung, Kibble, KulishFaddeev}
\begin{equation}
f_i(k)
=
e\sum_{a\in i}
\eta_a
\frac{p_a\cdot\varepsilon(k)}
     {p_a\cdot k},
\end{equation}
where the sum runs over the charged external particles
belonging to the hard configuration $i$, and
$\eta_a=\pm1$ distinguishes outgoing and incoming
charges. More generally, the construction of infrared-finite dressed states
is constrained by gauge invariance and asymptotic symmetries, and
appropriate dressings beyond the original Faddeev--Kulish prescription
have been discussed in Refs.~\cite{HiraiSugishita2019,
HiraiSugishita2021,HiraiSugishita2023}.
Since $p_a\cdot k \propto \omega$ in the soft limit,
the difference of two dressing profiles behaves as

\begin{equation}
|f_i(k)-f_j(k)|^2
=
\frac{{\cal B}_{ij}(\hat{\bm k})}{\omega^2},
\end{equation}
where ${\cal B}_{ij}(\hat{\mathbf k})$ collects the
angular dependence of the soft-photon dressing factors.
Using

\begin{equation}
\frac{d^3k}{2\omega_k}
=
\frac{\omega d\omega d\Omega}{2},
\end{equation}

one finds

\begin{equation}
\int_{\lambda}^{\omega_{\max}}
\frac{d^3k}{2\omega_k}
|f_i(k)-f_j(k)|^2
=
B_{ij}
\int_{\lambda}^{\omega_{\max}}
\frac{d\omega}{\omega},
\end{equation}

where

\begin{equation}
B_{ij}
=
\frac12
\int d\Omega
{\cal B}_{ij}(\hat{\bm k})
\end{equation}

is determined by the angular distribution of the soft
radiation.

The overlap therefore contains the characteristic
infrared logarithm

\begin{equation}
\int_{\lambda}^{\omega_{\max}}
\frac{d\omega}{\omega}
=
\ln\frac{\omega_{\max}}{\lambda}.
\end{equation}
Consequently, the magnitude of the overlap of the unresolved
soft-photon clouds is
\begin{equation}
\left|
{\cal D}_{ij}(\omega_{\max},\lambda)
\right|
\simeq
\exp\left[
-\frac{1}{2}B_{ij}
\ln\frac{\omega_{\max}}{\lambda}
\right].
\end{equation}
This quantity should be distinguished from the inclusive cross
section discussed in Sec.~II.  In the latter, real and virtual
infrared contributions combine so that the auxiliary regulator
$\lambda$ cancels, leaving the physical detector-resolution scale
$\omega_{\max}$.  Eq.~(17), by contrast, is the overlap of the
unresolved coherent soft clouds themselves.  Its infrared dependence
expresses the increasing distinguishability of different soft
records as a larger range of soft modes is traced over.

\section{Observable memory}

The relation between the soft-sector overlap and electromagnetic
memory can be made quantitative. To make this connection explicit,
let $i$ label a scattering branch from a common initial hard
configuration $\alpha$ to a final configuration $\beta_i$.
Following Hirai and Sugishita, the leading electromagnetic memory
associated with this branch is identified with the coefficient of
the leading soft factor $R_{\beta_i,\alpha}^A$. We therefore define,
in our normalization conventions,
\begin{equation}
{\cal M}_i^A(\hat{\bf k})
\equiv
\lim_{\omega\to0}
\omega R_{\beta_i,\alpha}^A(\omega,\hat{\bf k})
=
\sum_{a\in\alpha,\beta_i}
\eta_a e_a
\frac{
p_a\cdot\epsilon^A(\hat{\bf k})
}{
p_a\cdot(1,\hat{\bf k})
}.
\end{equation}
It follows from Eqs.~(12)--(15) that the coefficient $B_{ij}$
appearing in the soft-cloud overlap can be written as
\begin{equation}
B_{ij}
=
\frac{1}{2}
\int d\Omega
\sum_{A=1,2}
\left|
{\cal M}_i^A(\hat{\bf k})
-
{\cal M}_j^A(\hat{\bf k})
\right|^2 .
\end{equation}
Thus $B_{ij}$ is the angular norm of the difference between the
electromagnetic memory records associated with the two hard
configurations.  Substituting Eq.~(19) into Eq.~(17), the magnitude of the
resolution-dependent soft-cloud overlap can be written directly
in terms of the difference between the electromagnetic memory
records as

\begin{equation}
\begin{aligned}
\left|
{\cal D}_{ij}(\omega_{\max},\lambda)
\right|
\simeq {}&
\exp\Bigg[
-\frac{1}{4}
\ln\frac{\omega_{\max}}{\lambda}
\\
&\times
\int d\Omega
\sum_{A=1,2}
\left|
{\cal M}_i^A(\hat{\bf k})
-
{\cal M}_j^A(\hat{\bf k})
\right|^2
\Bigg].
\end{aligned}
\end{equation}

Although the absolute overlap in Eq.~(20) retains the auxiliary
infrared regulator $\lambda$, its change under a variation of the
detector resolution is regulator independent.  We therefore compare
the same off-diagonal coherence at two different detector resolutions
$\omega_1$ and $\omega_2$ and define the relative coherence loss by

\begin{equation}
{\cal I}_{ij}(\omega_2,\omega_1)
\equiv
-\ln
\left|
\frac{
{\cal D}_{ij}(\omega_2,\lambda)
}{
{\cal D}_{ij}(\omega_1,\lambda)
}
\right|.
\end{equation}

Using Eq.~(20), one obtains

\begin{equation}
\begin{aligned}
{\cal I}_{ij}(\omega_2,\omega_1)
={}&
\frac{1}{4}
\ln\frac{\omega_2}{\omega_1}
\\
&\times
\int d\Omega
\sum_{A=1,2}
\left|
{\cal M}_i^A(\hat{\bf k})
-
{\cal M}_j^A(\hat{\bf k})
\right|^2 .
\end{aligned}
\end{equation}

The auxiliary infrared regulator has dropped out of this relative
quantity.  It therefore isolates the physical dependence on detector
resolution.

It is convenient to introduce the squared distance between two
electromagnetic memory records,
\begin{equation}
d_{\rm mem}^2(i,j)
\equiv
\frac{1}{2}
\int d\Omega
\sum_{A=1,2}
\left|
{\cal M}_i^A(\hat{\bf k})
-
{\cal M}_j^A(\hat{\bf k})
\right|^2 .
\end{equation}
With the normalization of Eq.~(19), this is simply
$d_{\rm mem}^2(i,j)=B_{ij}$.

Equation~(22) then takes the compact form
\begin{equation}
{\cal I}_{ij}(\omega_2,\omega_1)
=
\frac{1}{2}
d_{\rm mem}^2(i,j)
\ln\frac{\omega_2}{\omega_1}.
\end{equation}
Equivalently, for a continuously varied detector resolution,
\begin{equation}
\boxed{
\frac{\partial {\cal I}_{ij}}
{\partial\ln\omega_{\max}}
=
\frac{1}{2}
d_{\rm mem}^2(i,j)
}
\end{equation}
This relation is the central operational consequence of the present
construction.  The logarithmic resolution flow of the hard-sector coherence obtained from the unresolved soft-cloud trace is fixed directly by the squared distance
between the electromagnetic memory records of the two hard
configurations.
Eq.~(25) describes the resolution flow obtained when the
unresolved soft clouds are traced in the form used above.  It is
important, however, to distinguish this result from the
superselection-compatible reduced description of properly dressed
asymptotic states.  In the generalized Chung construction of Hirai
and Sugishita, the final-state dressing can be decomposed as

\begin{equation}
D_{\beta}^A(k)
=
-R_{\beta}^A(k)
+\widetilde D_{\beta}^A(k)
+D^A(k),
\end{equation}

where the leading $1/\omega$ part $R_{\beta}^A$ carries the
electromagnetic memory, while
$\widetilde D_{\beta}^A=o(\omega^{-1})$ as $\omega\to0$.
Consequently, for two final branches $\beta_i$ and $\beta_j$, the
leading memory difference cancels from the properly dressed
interference factor,

\begin{equation}
R_{\beta_j}^A-R_{\beta_i}^A
+
D_{\beta_j}^A-D_{\beta_i}^A
=
\widetilde D_{\beta_j}^A
-
\widetilde D_{\beta_i}^A .
\end{equation}

This observation leads to a qualitatively different infrared
resolution flow.  Defining
$\Delta\widetilde D_{ij}^A
=\widetilde D_{\beta_j}^A-\widetilde D_{\beta_i}^A$,
the relative coherence loss between two detector resolutions is

\begin{equation}
{\cal I}_{ij}^{\rm dr}(\omega_2,\omega_1)
=
\frac{1}{2}
\int_{\omega_1<\omega<\omega_2}
\widetilde{d^3k}\,
\sum_A
\left|
\Delta\widetilde D_{ij}^A(k)
\right|^2 ,
\end{equation}

where
$\widetilde{d^3k}=d^3k/[(2\pi)^3 2\omega]$ in the normalization of
Hirai and Sugishita.  Hence

\begin{equation}
\Gamma_{ij}^{\rm dr}(\omega_{\max})
\equiv
\frac{\partial{\cal I}_{ij}^{\rm dr}}
{\partial\ln\omega_{\max}}
=
\frac{\omega_{\max}^{2}}
{4(2\pi)^3}
\int d\Omega
\sum_A
\left|
\Delta\widetilde D_{ij}^A
(\omega_{\max},\hat{\bf k})
\right|^2 .
\end{equation}

Since the infrared-safe dress condition requires
$\widetilde D_{\beta}^A=o(\omega^{-1})$, it follows that

\begin{equation}
\lim_{\omega_{\max}\to0}
\Gamma_{ij}^{\rm dr}(\omega_{\max})
=0 .
\end{equation}
Related suppression of deep-infrared radiation has been found in studies of subleading soft dressings.  Choi and Akhoury constructed asymptotic dressings consistent with subleading soft theorems \cite{ChoiAkhoury}, while Christodoulou and Toumbas showed explicitly that additional soft-photon emission below the characteristic dressing scale is strongly suppressed in dressed QED scattering \cite{ChristodoulouToumbas}.  The resolution-flow formulation considered here provides a complementary characterization in terms of the variation of hard-sector coherence with the detector resolution.

Eqs.~(25) and (30) therefore exhibit two distinct infrared
behaviors.  The unresolved soft-cloud overlap has a finite
logarithmic resolution flow determined by the squared memory
distance, whereas in the properly dressed, superselection-compatible
description the leading memory contribution is absorbed into the
asymptotic dressing and the infrared resolution flow vanishes.  The
resolution dependence thus provides a diagnostic of how
electromagnetic memory is encoded in the asymptotic state rather
than an additional source of infrared decoherence.

\section{Discussion and Conclusions}

The conventional infrared-safe cross-section formulation and the
density-matrix description of soft-photon information provide
complementary views of the infrared structure of QED.  The
Bloch--Nordsieck cancellation removes the unphysical infrared
regulator, while the physical detector-resolution scale
$\omega_{\max}$ remains.  As emphasized in the traditional
treatment of infrared QED \cite{LL}, this scale has a direct
operational meaning: it specifies the maximum energy of soft
photons that remain unresolved by the measurement.

In the present work we have examined how this finite-resolution
description is related to electromagnetic memory.  For the
unresolved soft-cloud overlap, the coefficient $B_{ij}$ can be
expressed as the squared angular distance between the corresponding
leading electromagnetic memory records.  Comparing the same
off-diagonal coherence at two different detector resolutions removes
the auxiliary infrared regulator and gives Eq.~(25): the logarithmic
resolution flow is determined by one half of the squared memory
distance.

An important qualification follows when the asymptotic states are
properly dressed.  In the generalized Chung construction of Hirai
and Sugishita, the leading $1/\omega$ memory contribution is absorbed
into the asymptotic dressing and cancels from the interference
factor.  The remaining resolution dependence is controlled by the
subleading dressing difference
$\Delta\widetilde D_{ij}$.
Since the infrared-safe dress condition requires
$\widetilde D=o(\omega^{-1})$, the corresponding resolution-flow
rate vanishes in the strict infrared limit, as expressed by
Eq.~(30).

The contrast between Eqs.~(25) and (30) is the main result of the
present analysis.  The unresolved soft-cloud description has
a finite logarithmic infrared flow proportional to the squared
memory distance, whereas the properly dressed,
superselection-compatible description has no leading infrared
flow.  Varying the detector resolution therefore provides an
operational diagnostic of how electromagnetic memory is encoded in
the asymptotic quantum state.

This distinction should not be interpreted as the generation of
stochastic classical infrared fluctuations or as fundamental
information loss.  The complete hard-plus-soft state remains
coherent.  Rather, the resolution dependence reflects how soft
information is partitioned between unresolved radiation and
asymptotic dressing.  In this way the finite-resolution viewpoint
of Berestetskii--Lifshitz--Pitaevskii \cite{LL} can be connected directly with
the modern dressed-state and electromagnetic-memory description of
infrared QED.

\end{document}